\documentclass[compsoc]{IEEEtran}
\usepackage{subcaption}
\usepackage{algorithm}
\usepackage{algorithmicx}
\usepackage{algpseudocode}
\usepackage{booktabs}
\usepackage[bottom]{footmisc}
\usepackage{amssymb}
\usepackage[nolist]{acronym}
\usepackage{amsmath}
\usepackage{amsfonts}
\usepackage{adjustbox}
\usepackage[english]{babel}
\usepackage{colortbl}
\usepackage{dirtree}
\usepackage{enumerate}
\usepackage[breaklinks]{hyperref}
\usepackage[utf8]{inputenc}
\usepackage{csquotes}
\usepackage{lscape}
\usepackage{longfigure}
\usepackage{longtable}
\usepackage{rotating}
\usepackage{multirow}
\usepackage{multicol}
\usepackage[numbers]{natbib}
\usepackage{xspace}
\usepackage{xcolor}
\usepackage{tcolorbox}
\usepackage{bm}

\definecolor{mygreen}{HTML}{61AB00}
\definecolor{myred}{HTML}{E74C3C}
\definecolor{myblue}{HTML}{005FBD}

\newcommand{\mypar}[1]{\smallskip\noindent\textbf{#1.}\xspace}

\newcommand{\system}{CyFence\xspace}

\DeclareCaptionType{mycapequ}[][List of equations]
\begin{document}

\title{\system: Securing Cyber-physical Controllers Via Trusted Execution Environment}

\author{Stefano~Longari,
        Alessandro~Pozone,
        Jessica Leoni,
        Mario~Polino,
        Michele~Carminati~\IEEEmembership{Member,~IEEE,},
        Mara~Tanelli,~\IEEEmembership{Senior Member,~IEEE,}
        and~Stefano~Zanero,~\IEEEmembership{Senior Member,~IEEE.}
\thanks{All authors are with the Dipartimento di Elettronica e Informazione, Politecnico di Milano, Italy. E-mail: \{stefano.longari, jessica.leoni, mario.polino, michele.carminati, mara.tanelli, stefano.zanero\}@polimi.it. alessandro.pozone@mail.polimi.it}
\thanks{Manuscript received October 30th, 202 2; revised X.}}

%
%
\begin{acronym}
	{\small
	\acro{CPS}{Cyber-physical System}
	\acro{CAN}{Controller Area Network}
	\acro{CAN-FD}{Controller Area Network with Flexible Data-rate}
	\acro{DSRC}{Dedicated Short Range Communications}
	\acro{ECU}{Electronic Control Unit}
	\acro{TCU}{Telematic Control Unit}
	\acro{OBD-II}{On-Board Diagnostic}
	\acro{OBD}{On-Board Diagnostic}
	\acro{IoT}{Internet of things}
	\acro{OEM}{Original Equipment Manufacturer}
	\acro{MOST}{Media Oriented System Transport}
	\acro{ADAS}{Advanced Driver Assistance System}
	\acro{ABS}{Antilock Braking System}
	\acro{LIN}{Local Interconnect Network}
	\acro{TPMS}{Tire Pressure Monitoring System}
	\acro{UDS}{Unified Diagnostic System}
	\acro{IDS}{Intrusion Detection System}
	\acro{ADS}{Anomaly Detection System}
	\acro{LSTM}{Long Short-Term Memory}
	\acro{OTA}{Over The Air}
	\acro{RNN}{Recurrent Neural Network}
	\acro{HMM}{Hidden Markov Model}
	\acro{ACK}{Acknowledgment}
	\acro{CRC}{Cyclic Redundancy Check}
	\acro{IFS}{Interframe Space}
	\acro{TEC}{Transmit Error Count}
	\acro{REC}{Receive Error Count}
	\acro{LTE}{Long Term Evolution}
	\acro{DoS}{Denial of Service}
	\acro{BC}{Bit Counter}
	\acro{PC}{Polarity Counter}
	\acro{SoF}{Start of Frame}
	\acro{DLC}{Data-Length Code}
	\acro{IDE}{Identifier Extension}
	\acro{RPM}{Repetitions Per Minute}
	\acro{RTR}{Remote Transmission Request}
	\acro{EoF}{End of Frame}
	\acro{HSM}{Hardware Security Module}
	\acro{IF}{Intermission Field}
	\acro{DL}{Deep Learning}
	\acro{ML}{Machine Learning}
	\acro{OCSVM}{One-Class SVM}
	\acro{CAN-H}{CAN High}
	\acro{CAN-L}{CAN Low}
	\acro{IRS}{Intrusion Reaction System}
	\acro{CRF}{Conditional Random Field}
	\acro{ROC}{Receiver Operating Characteristic Curve}
	\acro{AUC}{Area Under the Curve}
	\acro{MAC}{Message Authentication Code}
	\acro{GRU}{Gated Recurrent Unit}
	\acro{V2X}{Vehicle to Everything}
	\acro{FMS}{Fleet Management System}
	\acro{TCS}{Traction Control System}
	\acro{DBC}{CAN Data Base}
	\acro{FPR}{False Positive Rate}
	\acro{VAR}{Vector Auto Regression}
    \acro{NN}{Neural Network}
	\acro{ELU}{Exponential Linear Unit}
	\acro{MSE}{Mean Squared Error}
	\acro{TPR}{True Positive Ration}
	\acro{DR}{Detection Rate}
	\acro{MCC}{Matthews Correlation Coefficient}
	\acro{TTP}{Testing Time per Packet}
	\acro{AR}{Auto Regressive}
	\acro{FP}{False Positive}
	\acro{TEE}{Trusted Execution Environment}
	\acro{FCD}{Floating Car Data}
	\acro{GPS}{Global Positioning System}
	\acro{GNSS}{Global Navigation Satellite System}
	\acro{P2P}{Peer-to-Peer}
	\acro{PoW}{Proof-of-Work}
	\acro{PoS}{Proof-of-Stake}
	\acro{PBFT}{Practical Byzantine Fault Tolerance}
	\acro{FCC}{Federal Communications Commission}
	\acro{PRNG}{Pseudorandom Number Generator}
	\acro{MoST}{Monaco SUMO Traffic}
	\acro{GSM}{Global System for Mobile Communications}
	\acro{PoI}{Proof-of-Importance}
	\acro{PoB}{Proof-of-Believability}
	\acro{ITS}{Intelligent Transportation System}
	\acro{IVI}{In-Vehicle Infotainment}
	\acro{C-V2X}{Cellular V2X}
	\acro{RSU}{Road Side Unit}
	\acro{PKI}{Public Key Infrastructure}
	\acro{CDAL}{Controller Data Abstraction Layer}
	\acro{EMB}{Electro-Mechanical Braking system}
	\acro{PID}{Proportional Integral Derivative}
	\acro{DNN}{Deep Neural Network}
}
\end{acronym}

\IEEEtitleabstractindextext{%
%

\begin{abstract} 

In the last decades, \acp{CPS} have experienced a significant technological evolution and increased connectivity, at the cost of greater exposure to cyber-attacks. Since many CPS are used in safety-critical systems, such attacks entail high risks and potential safety harms. Although several defense strategies have been proposed, they rarely exploit the cyber-physical nature of the system. 

In this work, we exploit the nature of CPS by proposing \system, a novel architecture that improves the resilience of closed-loop control systems against cyber-attacks by adding a semantic check, used to confirm that the system is behaving as expected. To ensure the security of the semantic check code, we use the Trusted Execution Environment implemented by modern processors. 

We evaluate \system considering a real-world application, consisting of an active braking digital controller, demonstrating that it can mitigate different types of attacks with a negligible computation overhead.
\end{abstract}

}

\maketitle

\section{Introduction}
\label{sec:intro}

The technological progress of the last few decades has paved the way for the development and ubiquitous implementation of \acp{CPS}. They consist of computer-controlled devices capable of affecting the behavior of the referred physical system. Typical examples are industrial robots and vehicles. These devices are often safety-critical since their capabilities may harm humans if their tasks are performed incorrectly. For example, an automotive \ac{ABS} that fails in releasing the brake or a collaborative robot that moves unpredictably ~\cite{cardenas2011attacks}.
In the last few years, thanks to the development of cheap advanced communication technologies, we witnessed the implementation of wireless communication in many computer-controlled systems, allowing them to stream data from remote reliably. Moreover, worth mentioning are industry 4.0 robots and establishments, vehicle communication~\cite{DBLP:conf/ccs/TronLCPZ22}, and \ac{IoT} devices deployed in critical infrastructure. Indeed, this increased connectivity offers the opportunity to develop new features such as remote control and monitoring, making it particularly appealing for \acp{CPS}. However, from a security perspective, it also widens the attack surface of the system, exposing such devices to malicious attacks. Unfortunately, as of today, it is almost effortless to find ways to tamper with connected devices, and, in the literature, there are several documented vulnerabilities (e.g.,~\cite{quarta2017experimental1, miller2015remote}). 
It follows that the relentless implementation of \acp{CPS} in more and more applications increases the systems' safety but also their vulnerabilities. Therefore, it is of utmost importance to find new solutions to make these devices not only safe, as they are today, but also secure.
In the literature, several approaches have been presented, aiming at addressing the aforementioned issue of proposing secure \acp{CPS} \cite{ding2018survey}. However, the majority of solutions are borrowed from the embedded systems or \ac{IoT} security field, where \acp{HSM} are used to certify the cryptographic functions and communication, and secure boot ensures the authenticity of the executed firmware \cite{hassan2019current}. Preliminary studies have been conducted to address architectural approaches within the larger scale of entire \acp{CPS}~\cite{longari2019secure}. Also, researchers recently began to consider the use of new hardware technologies or software architectures to improve the security of the devices. An example is the Contego-TEE architecture~\cite{hasan2019protecting} which uses the ARM TrustZone \ac{TEE} to add a new security check on the final software computations~\cite{kim2018securing}.
However, in many control systems, like those employed in the automotive field, the problem is still open, since the spectrum of feasible solutions is strongly reduced by strict requirements and stringent deadlines to ensure real-time communication and actuation. This requirement is unsatisfied by most of the solutions proposed so far, which, unfortunately, are highly time-demanding. 

To overcome these issues, we propose \system, a novel software architecture that leverages a \ac{TEE} to perform a security check that can detect unexpected behaviors in a control loop-based system, hence focusing on the cyber-physical properties of \acp{CPS} such as industrial robots and machinery, vehicles, and critical infrastructures. 
In fact, we use control theory notions to develop a set of bounds for the system output that are both reliable and fast to compute. The proposed architecture is general, and it is designed to be compliant with most control algorithms. The main advantage of \system is related to the fact that it directly makes use of the \ac{TEE} technology of the micro-controller, without requiring additional hardware. 

In order to verify the effectiveness of the proposed architecture in a real-world case scenario, we implement a proof-of-concept~\footnote{The implementation is available at~\url{url.to.be.released}}, using as a case study the automotive domain, and specifically an \ac{ABS} system. Specifically, we evaluate the performance overhead of \system on an ARM Cortex-M architecture which implements TrustZone as \ac{TEE}. Our results indicate that the security check can be performed with minimal overhead and that the detection time of an attack is strictly correlated with the difference in magnitude between the tampered output and its correct value. 

\begin{figure*}
\centering
    \includegraphics[width=0.95\textwidth]{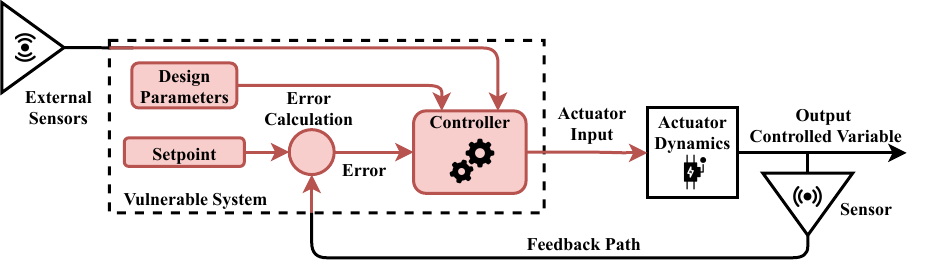}
    \caption{Standard CPS Closed-loop controller. In red, the elements that we consider in the threat model.}
    \label{fig:closedloopcontroller}
\end{figure*}

In detail, our contributions are:

\begin{itemize}
    
\item We present \system, a novel detection method that makes use of \acp{TEE} to secure closed-loop control systems by detecting anomalies modeled as deviations from their expected behavior. To the best of our knowledge, \system is the first solution that exploits \ac{TEE} technologies in the \acp{CPS} domain;

\item We demonstrate the practical feasibility of our approach by implementing \system on one the most pervasive \ac{TEE} technologies (ARM TrustZone) and studying its performance under real-time constraints;

\item We demonstrate the effectiveness of \system by considering the automotive \ac{ABS} system as a use-case scenario, showing its capabilities in preventing malicious behavior. 
    
\end{itemize}

The rest of the paper is structured as follows: in Section~\ref{sec:tzbackground} we present the \ac{TEE} technology along with a brief primer on closed-loop controllers. In Section~\ref{sec:related} we present the current state of the art regarding countermeasures for connected closed-loop control systems and the motivation of our work. Then, in Section~\ref{sec:threat} we detail our threat and attacker models. We proceed to describe our approach and the model of our architecture in Section~\ref{sec:tzapproach}, and the design and implementation of our case study, alongside our results, in Section~\ref{sec:tzcasestudy}. 
Finally, in Section~\ref{sec:limitation} we discuss the limitations and future works of \system, while we draw our conclusion in Section~\ref{sec:tzconclusion}.

\section{Background}
\label{sec:tzbackground}

To aid in comprehending our design, we provide two brief primers on closed-loop controllers and \ac{TEE} technologies.

\subsection{Closed-loop controllers} 

A system works in closed-loop when a feedback mechanism modifies the input based on information retrieved by the produced output in order to reduce errors and improve stability~\cite{visioli2006practical, aastrom2013computer}. In Figure~\ref{fig:closedloopcontroller}, we show a generic closed-loop controller. 
When the system aims at making the output reach and maintain a specific value, the so-called setpoint, it is a common practice to use as the input value the error signal computed as the difference between the state of the controlled variable (i.e., the signal transmitted through the feedback path) and the desired value (i.e., setpoint). This task is usually performed by a computation unit that subtracts the system output measured by a sensor (i.e., the feedback sensor) from the theoretical setpoint.
The controller leverages the previous inputs and outputs to compute 
the next input value required to reach the setpoint.

To design the controller for the application at hand, we choose a \ac{PID}, as it is the most widely adopted control-loop architecture, thanks to its simplicity and fast computation, making it an ideal choice for real-time applications. Although more advanced control approaches exist, the \ac{PID} is less computationally demanding, ensuring that high performance can be delivered without sacrificing real-time performance~\cite{gul2020review}. Moreover, the simple \ac{PID} architecture, characterized by only a few parameters to fine-tune (i.e., the gains), provided reliable and high-performance results. Within the context of ABS design, which we implemented as an explanatory use-case, \ac{PID} and other similar linear and time-invariant controller architectures have been successfully used in practice, see e.g.,~\cite{abc}.
In \ac{PID} controllers, the control variable is computed based on the tracking error as the sum of three components, the Proportional, Integral, and Derivative terms, respectively.
Equation~\eqref{eq:frequencypid} shows an ideal \ac{PID} transfer function in the frequency domain. 

\begin{tcolorbox}[boxrule=0.5pt]

\begin{equation} \label{eq:frequencypid}
C(s) = K_{p} + \frac{K_{i}}{s} + \frac{K_{d}s}{T_{f}s+1}
\end{equation}

\end{tcolorbox}

\subsection{Trusted Execution Environment Technologies}

\ac{TEE} is a technology that provides a secure execution environment for sensitive applications on a device. It is a hardware-based security technology that separates trusted and untrusted environments on a device, ensuring that sensitive data and processes are protected from tampering and other security threats. \ac{TEE} is implemented using a combination of hardware and software components, including a secure processor, secure boot, secure storage, and secure communication channels. The secure processor provides a trusted execution environment that is isolated from the main processor and the operating system. The secure boot ensures that only trusted software is loaded into the \ac{TEE}. The secure storage provides a secure area for storing sensitive data, such as cryptographic keys. The secure communication channels ensure that data transmitted between the TEE and the main processor is encrypted and protected from tampering. \ac{TEE} is widely used in mobile devices, \ac{IoT} devices, and other embedded systems to protect sensitive data, such as personal information, financial transactions, and cryptographic keys. Several TEE implementations are available today, including ARM TrustZone (and its variant Samsung TEEGRIS), Trustonic TEE, and Qualcomm Secure Execution Environment.
Even though our proposed solution can be implemented through a generic \ac{TEE} technology, in this work, we employ ARM TrustZone, given the widespread adoption of ARM-based processors in \ac{IoT} and embedded devices.

\mypar{ARM TrustZone} TrustZone~\cite{holding2009arm} enables two different software security states, \textit{secure} and \textit{non-secure}. It is possible to map which memory addresses can be accessed from the \textit{non-secure} state by creating two environments, each one allowing access to memory and executing code. Micro-controllers' peripherals are mapped as memory addresses and, therefore, can be treated as such and denied access from the \textit{non-secure} state if needed. The first executed code at startup is always from the \textit{secure} state. Finally, any attempted access to a \textit{secure} state from the \textit{non-secure} one without passing through the proper functions raises exceptions in the \textit{secure} state. To call a secure function from a \textit{non-secure} state, it is necessary to call an instruction in a memory map called ``non-secure callable'', which allows the transition. Therefore, ARM TrustZone provides only isolation between the \textit{secure} and \textit{non-secure} state. 
\section{Related Works}
\label{sec:related}

Several works proved the weaknesses of \ac{CPS} to cyber-attacks~\cite{finnicum2011building,amin2013cyber,liu2009false} since such systems have been connected to local and wide area networks, enabling remote access and increasing the attack surface. 

Various countermeasures have been proposed, the majority of which focus on the detection or the prevention of attacks at network or sensor level \cite{bhardwaj2019cyber,goh2017anomaly}. 

However, as stated by Giraldo et al. in an extensive survey on \ac{CPS} attack detection~\cite{giraldo2018survey}, "the physical evolution of the state of a system has to follow immutable laws of nature". Following this concept, researchers began designing novel detection solutions based on the physical properties of \ac{CPS}. Notable examples of such detection methods apply this concept to water control systems~\cite{hadziosmanovic2014through}, electricity consumption in smart meters~\cite{mashima2012research}, and boilers in power plants~\cite{wang2014srid}. 

The techniques applied to study the physical behavior of the system significantly vary from one work to the other.
Some techniques attempt to estimate the state of each sensor to detect those that are sending false information~\cite{chong2015observability, shoukry2020smt}. 
Other methods actively monitor the system by triggering events and analyzing the system response~\cite{mo2009secure, mo2015physical, valente2015using}. 
Other works attempt to detect whether the system executes the correct task by analyzing its power consumption~\cite{pu2020detecting}. 
Finally, some techniques use machine learning to implement detection~\cite{kiss2015clustering, hei2013pipac,longari2020cannolo}. 
For further details on detection systems and techniques for general \acp{CPS}, we refer the reader to Giraldo et al.'s work~\cite{giraldo2018survey}, and other surveys on the security of \acp{CPS}~\cite{altawy2016security,komninos2014survey,DBLP:journals/mam/YaacoubSNKCM20}.

For the purposes of this work, however, the most related works are those that also take into account the time constraints of real-time systems acting on \acp{CPS}.
Kim et al.\cite{kim2018securing} propose a security architecture that intends to virtually partition the memory space and enforce memory access control of a real-time device. With this approach, every process can only access its memory. 
Liu et al. \cite{liu2017protc} implement a trusted computing block within ARM TrustZone that enforces a secure access control policy for the essential protected peripherals of a drone. Aware of the several works proving the possibility to gain unrestricted access to all the peripherals, they propose an Access Control Decision Block that understands if the access to a peripheral is authorized and, eventually, allows its execution. While this paper focuses on securing drones, its rationale can be easily ported to other \ac{CPS}. 
Hasan et al. \cite{hasan2019protecting} propose an invariant checking mechanism to ensure the security and safety of the physical system. 
Yoon et al. \cite{yoon2017virtualdrone} design a software architecture that enables an attack-resilient control of modern UAS using virtualization (in which an unverifiable controller is supported by an attack resilient one). All the drone peripherals are virtualized, so the unverifiable controller cannot directly act on them. The safety controller takes over if any anomaly is detected.

\mypar{Discussion} One of the common traits of many state-of-the-art solutions is their being heavily context-dependent, making them hardly extensible to other systems. 
Instead, our approach positions itself as a context-independent security architecture capable of detecting tampering at a controller level. In fact, it can be used with the vast majority of closed-loop control systems since it is control-algorithm agnostic.
Contrary to computation-heavy approaches such as those based on machine learning, \system directly exploits the physical behavior of the system through a set of semantic boundaries. This solution has a lower impact on the controller performances guaranteeing it can work on \ac{CPS} with stringent time constraints (e.g., an automotive \ac{ABS} such as the one presented in our case study). 

\section{Threat Model}
\label{sec:threat}

In the context of \acp{CPS}, threat modeling is fundamental to comprehend the risks associated with the system and to assess its security. 
In Figure~\ref{fig:closedloopcontroller} we show, colored in red, the attack surface of our threat model.   
In this specific context, we consider an attacker that modifies the controller code and parameters to compromise the availability or the integrity of the CPS by performing either a \textbf{\ac{DoS}} attack or a \textbf{Tampering} attack, where the final objective is to modify the output of the controller, consequently affecting also the actuator behavior.  
This can be achieved through either uploading tampered firmware on the controller unit, or by exploiting vulnerabilities that enable the attacker to execute arbitrary code, as demonstrated by Quarta et al.~\cite{quarta2017experimental1}.
Note that, even assuming that security measures such as secure boot or secure software update are implemented, the threat model presented above remains valid, since such measures are ineffective against the exploitation, at run time, of memory corruption vulnerabilities.
Moreover, the intuition of exploiting \acp{TEE} to secure sensor data and detection algorithms enables \system to ensure that the attacker has no access to the security module while keeping such module on the same computing unit of the closed-loop controller, differently from other solutions such as \acp{HSM}. 

Although \system is independent of the performed attack, we assume that an attacker can tamper the actuator output either by tampering with the actuator input (i.e., produced in output by the controller) or by modifying the controller source code and parameters, which are stored in the device's memory. 
Specifically, we envision three possible attacks that an adversary can implement: 
\begin{enumerate}[(a)]
    \item The tampering of the controller parameters. In the case of a \ac{PID} controller, $K_{p}$, $K_{i}$, and $K_{d}$, presented in Equation~\ref{eq:frequencypid};
    \item The tampering of the controller setpoint;
    \item The tampering of the controller output. 
\end{enumerate}

\system is a software-only solution that exploits the capabilities of the \ac{CPS} micro-controller. Therefore, we consider the security of the external network, although of paramount importance, as out of the scope of this work and assume that state-of-the-art solutions are deployed to guarantee the network signal integrity. For this reason, we consider the information arriving from external sensors through networks -- such as the \ac{CAN} in vehicles -- as trusted. 

\begin{figure*}
\centering
    \includegraphics[width=0.9\textwidth]{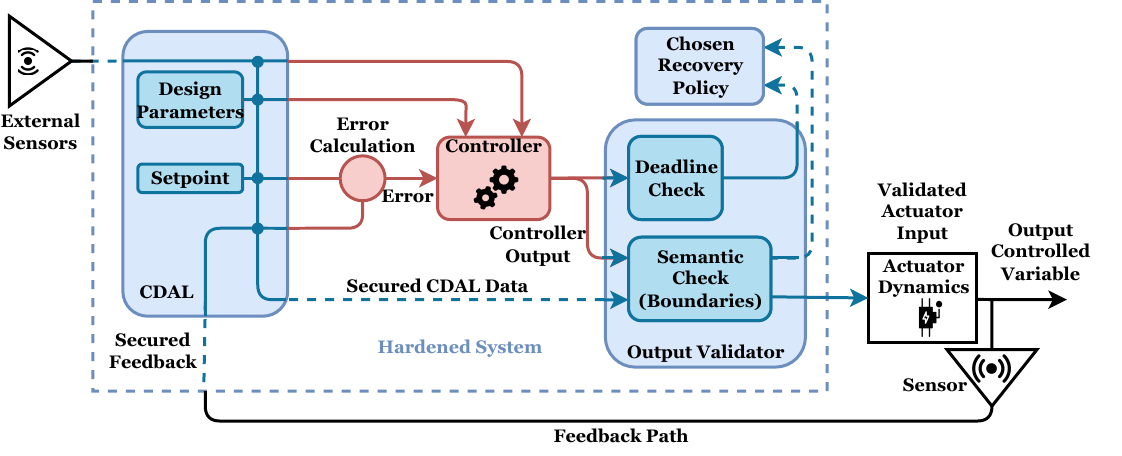}
    \caption{ \system's model. Components in blue and red are implemented in the secure state and non-secure state, respectively.}
    \label{fig:ProposedArchitecture}
\end{figure*}

\section{\system}
\label{sec:tzapproach}

We envision \system as a software architecture that guarantees the integrity of security measures such as intrusion and anomaly detection systems directly implemented in the CPS control unit. \system models the properties of the physical system in order to detect misbehavior of the controller by analyzing its outputs in relation to its parameters and inputs. By verifying that the outputs of the CPS conform to its expected physical behavior, \system can provide a solution that is independent of the specific implementation of the controller.

We exploit the \ac{TEE} to isolate and secure the inputs of the system and the output validation from the controller code to guarantee their integrity. This ensures that even if an attacker manages to compromise the controller, the isolated check would not be affected and could detect the attack. Starting from the threat model and attack surfaces presented in Section~\ref{sec:threat}, the idea behind \system is that of designing an integrity and semantic check of the controller output (i.e., an assessment of the compliance of the current behavior with the expected one). To do so, we first compute the expected system functioning through the control-theory-based model of the system under normal behavior and, through this model, we define a set of bounds within which we consider the controller output valid. We define these bounds as "semantic", since they are strictly dependent on the specific function of the modeled system. 

Figure~\ref{fig:ProposedArchitecture} shows a schematic view of \system architecture. Our solution comprises three different software components implemented in the secure state: the \textit{\ac{CDAL}}, which safely stores the inputs and parameters of the controller; the \textit{Output Validator}, which verifies that the outputs of the CPS conform to its expected physical behavior; the \textit{Recovery Policy}, which implements the reaction of the system upon misbehavior detection. These three components encompass the non-secured controller. 

\mypar{Explanatory Example} Considering a robotic arm CPS, an attacker may be able
to exploit a vulnerability (as explained in \cite{quarta2017experimental1}) and obtain code execution compromising the controller. In line with the threat model presented in \ref{sec:threat}, the attacker aims to alter the robotic arms' behavior by modifying the controller output. 
\system guarantees the attack detection and reaction. 
The attacker may attempt to modify the controller parameters or inputs to achieve its target output while hiding its presence. However, \system would detect any tampering of such variables since its CDAL, implemented in TEE, is the first to receive signal inputs and securely stores the controller parameters. 
The attacker may directly tamper with the controller code. Again, \system would detect such an attack since the output of the controller would not be coherent with the expected behavior modeled in the output validator. 

\mypar{TEE design choices} \ac{TEE} technology merely provides isolation between the \textit{secure} and the \textit{non-secure} state. Therefore, it is necessary to define which elements of the system are trusted (i.e., harmless and known to be secure) and map only those elements in the secure state.
Mapping the whole controller code to the secure state, besides making the solution controller-dependent, would lower the security of the entire system, increasing its attack surface and making the \ac{TEE} potentially useless. In fact, while it is generally possible to implement the entire control system in the \ac{TEE}, this would make the trusted element, implemented in the secure state, more complex and hence harder to verify. This contrasts with the general goal of having simple, easily verifiable trusted elements. This would be particularly true in cases where the controller is implemented through machine learning techniques as proposed in~\cite{zhang2000adaptive}. In addition, the control code may require updates, either for security or functionality reasons. An update procedure that modifies the memory in the secure state would increase the attack surface of the system. Moreover, it is also necessary to design a program control flow that does not allow an attacker to bypass or tamper with the secure code execution, e.g., by modifying inputs and outputs or hijacking the execution to skip the output validation. 
In our solution, to maintain trusted elements to a minimum, we isolate and secure only inputs and parameters (CDAL), the output validator, and the recovery policy.  

\mypar{Computational requirements} It is important to highlight that \system's requirements on execution time are bound to the refresh rate of the CPS under analysis. In fact, \system must be sufficiently fast to be executed at every control-loop iteration without introducing overheads that can exceed the real-time requirements of the system under analysis, hence causing a deadline miss. 

\subsection{\acf{CDAL}} 

This software component, at every control loop iteration, reads the  sensor output and extrapolates the values that are relevant to the controller. As mentioned above, sensor data can be mapped to allow access only from the secure state, ensuring their integrity as long as the network is considered secure. The \textit{\ac{CDAL}} also serves the purpose of secure memory for all the parameters computed at design time and required by the \textit{Output Validator} (the second software component). The goal of this software layer is that of storing and preparing in a secure environment the data used by the non-secured controller and the \textit{Output Validator} for their scopes. 

\subsection{Output Validator} 
\label{ssec:ov}

This software component checks the output of the controller and is further divided into two modules, the \textit{Deadline Check} and the \textit{Semantic Check}.

\begin{itemize}

\item The \textit{Deadline Check} It ensures that the controller does not miss the delivery of its computation results to the actuator. If the deadline is missed, a recovery policy is initiated.

\item The \textit{Semantic Check} has the task of guaranteeing that the produced output fulfills a series of requirements given by the \textit{apriori} knowledge of the controller structure.

\end{itemize}

Given a linear and time-invariant control loop, as that considered herein, with stability guarantees and without zeroes at the origin in the loop transfer function, the output movement is guaranteed to be enveloped by two exponential functions a priori known since dependent on the system transfer function. 
We use this knowledge to detect anomalous behaviors, identified by monitoring that the output does not exceed the theoretical bounds during the operations.
To do this, we approximate the closed-loop behavior using commonly employed models of the complementary sensitivity function~\cite{aastrom2010feedback}. Such approximation leads, in general, to a second-order transfer function, and the natural frequency and damping of its poles can be computed from the phase margin and the cross-over frequency. Thus, the bounds are computed as shown in Equation~\eqref{eq:bounds}.

\begin{tcolorbox}[boxrule=0.5pt]

\begin{equation}
bounds(t) = setpoint \pm e^{-\omega_{n}\xi t}
\label{eq:bounds}
\end{equation}

where $\omega_n$ is the cross-over natural frequency of the loop transfer function, and $\xi\approx \phi_m/100$, $\phi_m$ being the phase margin associated to the loop transfer function itself.

\end{tcolorbox}

There are two main reasons that make these bounds appealing for our application:
\begin{itemize}
    \item As long as the output value of the controller is within the bounds, the system evolves according to its expected behavior, eventually reaching the setpoint, hence maintaining its functionalities. 
    Accordingly, as long as the output remains within the boundaries we consider the \ac{CPS} not tampered.

    \item The bounds are computed using only parameters that are available at design time. This means that it is possible for the model to be stored completely in the secure state, making it impossible for the attacker to modify it. 
\end{itemize}

As a positive side note, from an implementation standpoint, the execution of the boundaries check is simple and fast, increasing the possibility for the system to comply with a real-time application. Even considering controllers with more complex boundary functions, it is possible to revert to pre-built lookup tables to minimize the computation overhead. 
Clearly, the computation of the bounds here is made in a simplified way, that does not take into account the model uncertainties that may affect the system dynamics. Note, however, that robust control procedures exist to compute such bounds also for uncertain systems so that the proposed procedure has a wide range of applicability.

\subsection{Recovery Policy} 
\label{ssec:rp}

\begin{figure}
    \begin{subfigure}{\linewidth}
        \centering
        \includegraphics[width=.95\linewidth]{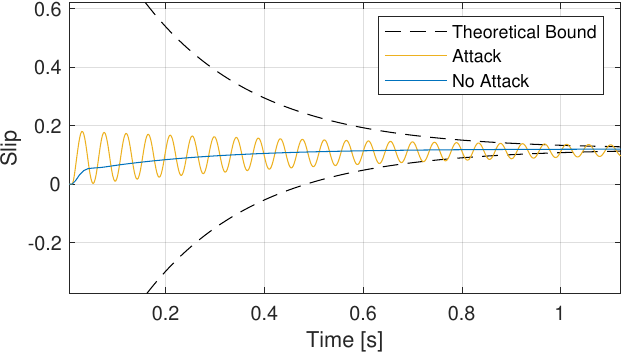}
        \caption{Modified $K_p$ attack example. The $K_{p}$ for this attack has been modified to 18,000.}
        \label{fig:KPAttackUnstable}
    \end{subfigure}    
    \begin{subfigure}{\linewidth}
        \centering
        \includegraphics[width=.95\linewidth]{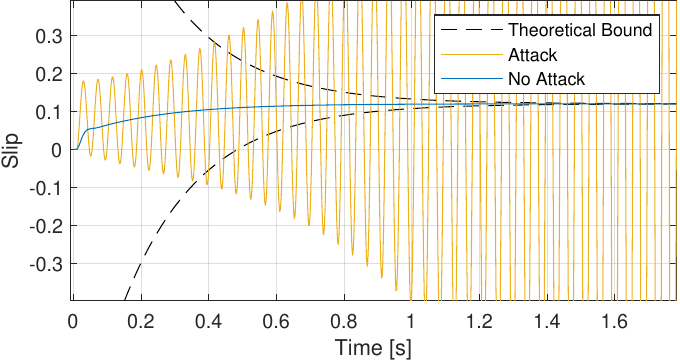}
        \caption{Modified $K_d$ attack example. The $K_{d}$ for this attack has been modified to 1,700.}
        \label{fig:KDAttackUnstable}
    \end{subfigure}
    
    \begin{subfigure}{\linewidth}
        \centering
        \includegraphics[width=.95\linewidth]{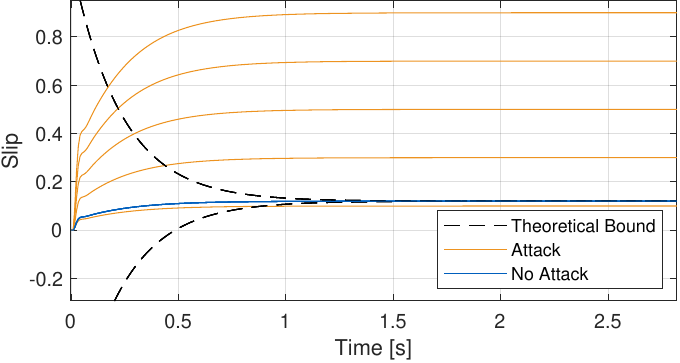}
        \centering
        \caption{Modified setpoint attack examples. Orange lines represent attacks that modify the setpoint from 0.1 to 1.3 with 0.3 steps.}
        \label{fig:FakeSetpoint}
    \end{subfigure}
    
    \begin{subfigure}{\linewidth}
        \centering
        \includegraphics[width=\linewidth]{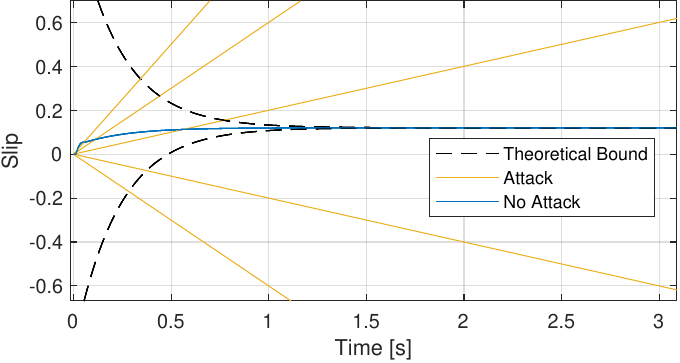}
        \centering
        \caption{Modified output attack examples. Orange lines represent attacks that modify the output from -0.6 to 1 with 0.4 steps.}
        \label{fig:FakeInput}
    \end{subfigure}
    \caption{Effects of attacks on the slip of the wheels when bounds are crossed.}
    \label{fig:attacks}
\end{figure}

The \textit{Recovery Policy} represents the sequence of actions taken by our system once an anomaly is detected. In particular, as mentioned before, the controller output must be comprised inside specific boundaries, computable considering its mathematical model. Therefore, if they are exceeded, the recovery policy is initiated. As an example, we can see in Figure~\ref{fig:KPAttackUnstable} how the boundaries (dotted black lines) are exceeded by the actual controller output (in orange) in the case where an attacker has modified the $K_p$ parameter of a \ac{PID} controller. The recovery policy would be triggered at $t\approx 0.4$ seconds, where the bounds are surpassed for the first time. 

As with any intrusion reaction system, the recovery policy is extremely context-dependent, and it is impossible to define a priori the perfect solution for every system to protect. Nonetheless, we suggest that a generally valid solution is that of implementing module redundancy, which implies having a second controller to activate in case the first one fails. Module redundancy is a standard in safety systems, adopted as a solution to provide error correction and not only error detection. 

To achieve this goal, at least three controllers are required, to allow for a voting mechanism to establish which is the faulty one. Our general recovery policy architecture suggestion entails designing two types of controllers, one connected, and one not connected. The first one, more complex, is updatable. It has more capabilities and is the one usually leveraged to control the actuator. The second one, instead, is designed \textit{ad-hoc} not to be reachable by any attacker and is only implemented as the recovery policy of the system. Please notice that this recovery policy is particularly effective in deterministic systems, such as the automotive \ac{ABS}, since the measure to undertake in order to ensure security is the same over time, \textit{i.e.,} stop the car. In more complex systems, which require more frequent task updates, this recovery policy would be harder to implement. However, even considering complex systems, such as cooperative robots, in which the optimal countermeasure may vary over time, our detection system can still be effective, setting as recovery policy a system shutdown.
\section{Case study: automotive Anti-lock Braking System}
\label{sec:tzcasestudy}

To study the feasibility of our approach and assess its performance, we implement it on an automotive \ac{ABS}~\footnotemark[1]. A more complex control system, such as those used in critical infrastructure or industrial machinery, may burden the readability and understandability of our approach. 
Instead, the ABS allows us to easily show the potential of \system, and easily understand the consequences and effects of attacks in a common and known environment. Nonetheless, our approach can be extended to any closed-loop control system modelable with a control scheme that works with a given setpoint and within an output-error minimization framework, both being elements of almost all closed-loop controller schemes.

To achieve this goal, we use two dynamical models as a starting point for controller design, one for modeling the vehicle dynamics on which the \ac{ABS} acts and one for the braking system (actuator). To model the longitudinal dynamics of the vehicle, we employ a simplified single-corner model, which considers a quarter of the vehicle and thus a single wheel, and we assume that the controlled variable used as output is the wheel slip $\lambda$. Such a variable represents the amount of skidding of the wheel, which, to maximize the braking force on the ground, must be kept as close as possible to the optimal working condition, known a priori and set as the setpoint. Using the data and model proposed in~\cite{tanelli2007combining}, the transfer function representing the dynamics from braking torque to wheel slip we get the model in Equation~\eqref{eq:transferfunctionsinglecorner}. 
For the actuator, we use a servo-controlled \acf{EMB} proposed again in~\cite{tanelli2007combining}, whose input/output behavior is described through the transfer function in Equation~\eqref{eq:transferfunctionbrake}.

The overall system dynamic to be controlled is hence presented in Equation~\eqref{eq:controlled_dynamic}. To apply our approach, we must comply with two requirements: the transfer function of the closed-loop system must be asymptotically stable and without zeroes at the origin. If this is fulfilled, the output boundaries given in Equation~\eqref{eq:bounds} can be defined \textit{a priori}. 
Therefore, we resort to one of the most common controller architectures, a \ac{PID}, and specifically that represented by Equation~\eqref{eq:controller}. Accordingly, the loop transfer function is represented in Equation~\eqref{eq:complete_dynamic}. 

\begin{tcolorbox}[boxrule=0.5pt]

\begin{equation}
G_{\lambda}(s) = \frac{\frac{r}{J\bar{v}}}{s+[\mu_{1}(\bar{\lambda})\frac{F_{z}}{m\bar{v}}(1-\bar{\lambda}+\frac{mr^{2}}{J} )]}   
\label{eq:transferfunctionsinglecorner}
\end{equation}

Transfer function of the single-corner model: $r$ is the wheel radius, $J$ is the moment of inertia, $v$ is the longitudinal speed of the vehicle, $\mu_{1}$ is the friction coefficient, $\bar{\lambda}$ is the longitudinal slip, m is the mass of the single corner model (1/4 of the vehicle mass), and $F_{z}$ is the normal force.

\begin{equation}
G_{caliper}(s) = \frac{\omega_{act}}{s+\omega_{act}}e^{s\tau}
\label{eq:transferfunctionbrake}
\end{equation}

Transfer function of the \ac{EMB}: $\omega_{act}$, the actuator bandwidth, is a measure of the system speed whose value for our model is 70$\frac{rad}{s}$, while $\tau$ represents the delay of the actuator, is due to both the actuator design and signal transmission and in our model is equal to 10$ms$.

\begin{equation}
D(s) = G_{\lambda}(s)G_{caliper}(s)
\label{eq:controlled_dynamic}
\end{equation}

\begin{equation}
R_{\lambda}(s) = K_{p} + \frac{K_{i}}{s} + \frac{K_{d}s}{T_{f}s+1}
\label{eq:controller}
\end{equation}

As also shown in the first column of Table~\ref{table:detectiontimes2}, the \textbf{nominal values} of the parameters used for the implementation of the \ac{ABS} controller, calculated through MATLAB's PID Tuner, are: $K_p = 3,151$, $K_i = 40,400$, $K_d = 30.5$, $Setpoint = 0.12$, and $T_{f} = 0.1$.

\begin{equation}
L(s) = R_{\lambda}(s)G_{\lambda}(s)G_{caliper}(s)
\label{eq:complete_dynamic}
\end{equation}

\end{tcolorbox}

To control the wheel slip, one needs to estimate the vehicle speed which, together with the wheel one, is needed to compute the value of this variable. In this work, we assume that the vehicle speed is estimated as presented in~\cite{tanelli2007combining}, and the estimator is implemented in the employed proof-of-concept simulator.

\begin{figure}
    \begin{subfigure}{\linewidth}
        \centering
        \includegraphics[width=.95\linewidth]{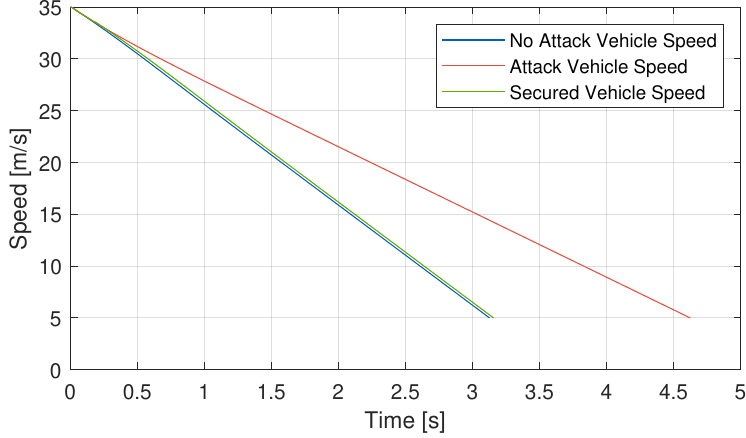}
        \caption{Vehicle speed in the braking simulation.}
        \label{fig:vspeed}
    \end{subfigure}    
    \begin{subfigure}{\linewidth}
        \centering
        \includegraphics[width=.95\linewidth]{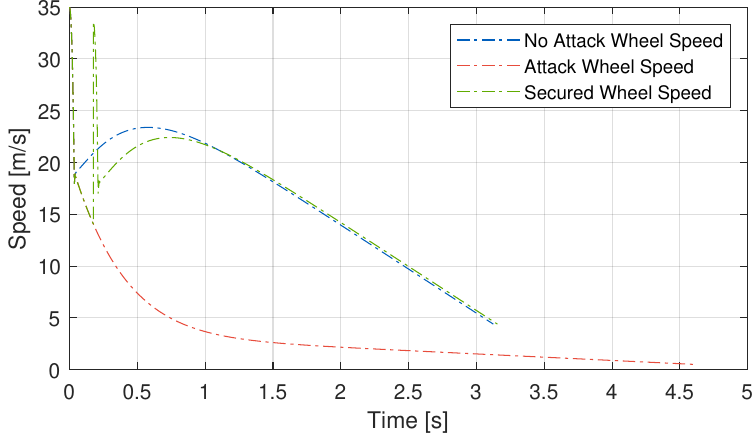}
        \caption{Wheel speed in the braking simulation.}
        \label{fig:wspeed}
    \end{subfigure}
    
    \begin{subfigure}{\linewidth}
        \centering
        \includegraphics[width=.95\linewidth]{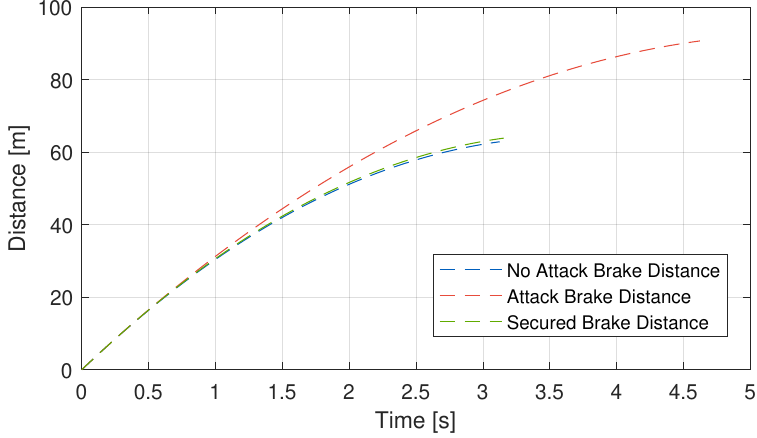}
        \centering
        \caption{Braking distance in the braking simulation.}
        \label{fig:bdistance}
    \end{subfigure}
    \caption{Behavior of a vehicle in terms of speed, wheel speed, and braking distance when engaging brakes (with a starting speed of 35 m/s to a final of 5 m/s). In blue the scenario where the ABS is untampered, in red the scenario where the ABS is under a $Setpoint = 0.9$ attack, and in green a scenario where the previous attack is performed, but \system is implemented.}
    \label{fig:attack_image}
\end{figure}

\subsection{Experimental Setup}

We implemented the full control system in Matlab/Simulink\textregistered, testing it on braking maneuvers carried out at different speed values and computing phase margins and gain crossover frequencies (i.e., the bounds) for all feasible scenarios.
We also implemented our proof of concept on a Nucleo-L552ZE-Q board with a Cortex-M33 micro-controller that supports TrustZone technology~\footnote{https://developer.arm.com/ip-products/security-ip/trustzone} to study the delays added by the context switches between secure and non-secure states. 

\begin{table}
    \centering
        \caption{Detection times for the various types of tested attacks for different parameter values.}
    \begin{tabular}{|c|cc|c|}
        \hline

         \textbf{Fixed Param.} & \multicolumn{2}{c|}{\textbf{Modified Param.}} & \textbf{Detection Time} \\
         \hline
         \hline
         \multirow{5}{*}{\shortstack[c]{ $K_i = 40,400$ \\ $K_d = 30.5$ \\ $Setpoint = 0.12$}} 
         &
         \multirow{5}{*}{\shortstack[c]{ $K_p$}} 
         &   18.000 &     0.720     s\\
         & & 18,500 & 0.539     s\\
         & & 19,000  & 0.406    s\\
         & & 19,500 & 0.356     s\\
         & & 20,000 & 0.311     s\\
         \hline
         \hline
         \multirow{5}{*}{\shortstack[c]{ $K_p = 3,151$ \\ $K_d = 30.5$ \\ $Setpoint = 0.12$}} 
         &
         \multirow{5}{*}{\shortstack[c]{ $K_i$}} 
         & 750,000 &  0.512 s\\
         & & 800,000 & 0.343    s\\
         & & 850,000 & 0.289     s\\
         & & 900,000 & 0.245    s\\
         & & 950,000 & 0.204     s\\
         \hline
         \hline
         \multirow{5}{*}{\shortstack[c]{ $K_p = 3,151$ \\ $K_i = 40,400$  \\ $Setpoint = 0.12$}} 
         &
         \multirow{5}{*}{\shortstack[c]{ $K_d$}} 
         &   1,600 &   0.615   s\\
         & & 1,650 &   0.480  s\\
         & & 1,700 &   0.392      s\\
         & & 1,750 &   0.308  s\\
         & & 1,800 &   0.301      s\\
         \hline
         \hline
         \multirow{5}{*}{\shortstack[c]{ $K_p = 3,151$ \\ $K_i = 40,400$ \\ $K_d = 30.5$}} 
         &
         \multirow{5}{*}{\shortstack[c]{ $Setpoint$}} 
         &   0.1 &  \textbf{0.880  s}\\
         & & 0.3 & 0.444  s\\
         & & 0.5 &  0.294  s\\
         & & 0.7 & 0.226 s\\
         & & 0.9 & \textbf{0.176  s}\\
         \hline
         \hline
         \multirow{5}{*}{\shortstack[c]{ $K_p = 3,151$ \\ $K_i = 40,400$ \\ $K_d = 30.5$ \\ $Setpoint = 0.12$}} 
         &
         \multirow{5}{*}{\shortstack[c]{ $Output$}} 
          & -0.6 & 0.285      s\\
         & & -0.2 & 0.377     s\\
         & & 0.2 & 0.771     s\\
         & & 0.6 & 0.445     s\\
         & & 1 & 0.344     s\\
         \hline
    \end{tabular}

    \label{table:detectiontimes2}
\end{table}

\subsection{Feasibility of detection and detection time}
\label{ssec:feasibility_detetction}

As mentioned in Section~\ref{sec:threat},  the attack implementation does not change the behavior of \system. However, for the sake of clarity, we simulate different attacks through the tampering of the setpoint, the parameters, or the output of the controller, as presented in Figure~\ref{fig:attacks}.
In all three considered scenarios, whether the attack is detected or not depends on whether the tampered controller output crosses the boundaries defined by \system as explained in Section~\ref{ssec:ov}. It follows that if an attacker modifies the controller's behavior without making the tampered variable exceed the defined boundaries, the attack will not be detected. 
However, this is not an issue as the controller guarantees that if the variable (i.e., the measured slip) is inside the theoretical boundaries, the closed-loop system remains asymptotically stable, and the \ac{ABS} works appropriately. Hence, until the tampering does not make the output exceed the boundaries, no countermeasure is needed. 
Besides assessing our system's effectiveness, when analyzing its performances, we are interested in analyzing the time required to detect the attack and its effects in situations where the output exceeds the theoretical boundaries (i.e., when the \ac{ABS} ceases to work properly). Although this is surely parameter-dependent, we refer the reader to the graphs shown in Figures~\ref{fig:KPAttackUnstable},~\ref{fig:KDAttackUnstable},~\ref{fig:FakeSetpoint}, and ~\ref{fig:FakeInput} that show examples of the \ac{ABS} behavior when attacks are deployed. In particular, we present the nominal behavior over time of the wheel slip in blue, the modeled boundaries computed by \system in dashed black, and the wheel slip in case of attack in orange. Misbehavior is detected when the slip exceeds the boundaries.
While the primary goal of Figure~\ref{fig:attacks} is that of showing different system's behaviors upon different degrees of perturbation on the variables under attack (i.e., parameters, setpoint, and output), it is important to highlight how larger perturbations have a greater impact on the system's behavior and, consequently, are detected earlier. 
This is further confirmed by the results in Table~\ref{table:detectiontimes2}, where we also show the actual results of multiple detection tests for all the three attacks we proposed. In the worst case (i.e., modified setpoint with setpoint = $0.1$),  the anomaly was detected in $0.88$ seconds. This is because, as shown in Figure~\ref{fig:FakeSetpoint}, the attacker's behavior is very similar to the unmodified behavior of the system when no attacks are implemented. Moreover, as further discussed in the next paragraph, the behavior of the \ac{ABS} is minimally affected by the attack.

\subsection{Braking simulation}
As an additional analysis of the effects of the attacks and the proposed architecture on the \ac{ABS}, we perform a simulation of the device behavior while a vehicle engages the brake action to a complete halt. 
The simulation starts with the vehicle moving at $35$ m/s ($126$ km/h) and stops when the vehicle reaches a speed of $5$m/s, at which the \ac{ABS} is deactivated by design. 

As a recovery policy, we implemented the module redundancy procedure explained in Section~\ref{ssec:rp}, switching to the second (unmodified) controller as soon as the system detects the attack and the main one can not be considered reliable. We investigate system performances considering the three modified Setpoint attacks, $Setpoint = 0.1$, $Setpoint = 0.5$, and $Setpoint = 0.9$, that represent the cases with the worst, average, and best detection times in the Exp.~\ref{ssec:feasibility_detetction}. 

Table~\ref{tab:SimulTimes} shows that if no attack is performed (i.e., nominal with $Sp=0.12$), the vehicle stops in $62.87$ meters. 
Considering the first attack scenario, i.e., setpoint to $0.1$, without any detection system, the braking distance increases by $0.76$ meters. Our approach reduces it to $0.69$ meters. This minor improvement is due to the fact that, as anticipated before, this attack has little to no effect on the actual braking behavior of the vehicle. When dealing with the $Setpoint = 0.9$ attack, the effectiveness of our detection method is more evident: where the attack would increase the braking space by $27.81$ meters, once the anomaly is detected (after $0.176$ seconds), the correct \ac{ABS} functionality is reinstated, limiting the increase in braking space to only $0.99$ meters. 

In Figure~\ref{fig:attack_image}, we graphically present the effects of \system on the braking distance, the speed of the vehicle, and the speed of the wheels under the $Setpoint = 0.9$ attack. We present the untampered behavior of the vehicle in blue, the behavior of the vehicle under attack in red, and the behavior of the vehicle under attack with \system implemented in green. It is evident how the attack increases the braking time (and space) of \~1.5 seconds (and \~27 meters). However, if \system is implemented after it detects the attack, it reduces the braking space to less than a meter more than the nominal behavior. Moreover, without \system, the wheel speed~\ref{fig:wspeed} under attack is significantly lower than the speed of the vehicle since the wheels are slipping.

\begin{table}
\centering
\caption{Comparison of the stopping distances of the \ac{ABS} with and without \system's detection and recovery policy.}
\label{tab:SimulTimes}
\begin{tabular}{|c|c|c|c|}
\hline
\multirow{2}{*}{Parameter} & Stop distance for & Detection & Stop distance \\ 
& non-secure device & Time & for secure device\\ \hline \hline
\textbf{Nominal} & $62.87m$ & --  & $62.87m$\\ \hline
$Sp = 0.1$ & $63.63m$ (\textbf{$+0.76m$}) & $0.88s$ & $63.56$  (\textbf{$+0.69m$})\\ \hline
$Sp = 0.5$ & $71.87$  (\textbf{$+8.99m$}) & $0.29s$ & $64.20$ (\textbf{$+1.32m$})\\ \hline
$Sp = 0.9$ & $90.68$  (\textbf{$+27.81m$}) & $0.18s$ & $63.87$ (\textbf{$+0.99m$})\\ \hline
\end{tabular}
\end{table}

\subsection{Computational performances analysis}

We estimate the overhead of \system on a standard \ac{ABS} implementation for each controller loop iteration by measuring the time required to execute the added code (for the bounds generation and semantic check) and for transitioning between secure and non-secure states. We made such experiments on a Nucleo-L552ZE-Q board~\footnote{https://www.st.com/en/evaluation-tools/nucleo-l552ze-q.html}, which is a low-medium end TrustZone equipped development board, the results are also reported in Table~\ref{table:timings}.

The semantic check and bounds generation is the primary source of added delay, and depending on the implementation, either through a \textit{math.h} function or through a lookup table, it causes a timing overhead of $490\mu$s and $15\mu$s, respectively.
The transitioning from secure to non-secure state and vice-versa cost respectively $8\mu$s and $4\mu$s.

The total overhead required by our solution is the sum of the two transitions with the semantic check and bounds generation, consequently, depending on the implementation of the latter, we obtain $502\mu$s or $27\mu$s respectively.

In the case of a typical implementation of an automotive \ac{ABS} the loop frequency is usually $200$Hz, which requires a deadline of $5$ms for the computation of the whole loop. The overall cost for both implementations is therefore negligible considering our application requirements, making our approach feasible and effective. 

\begin{table}[htb]
\centering
\caption{Timings of \system functions on a Nucleo-L552ZE.}
\label{table:timings}
\begin{tabular}{|c|c|}
\hline
\textbf{Function} & \textbf{Time} \\ \hline \hline

 Semantic Check (math.h)  & $490\mu$ \\ \hline
 Semantic Check (LUT)  & $15\mu$ \\ \hline
 Secure to Non-Secure  & $8\mu$ \\ \hline
 Non-Secure to Secure & $4\mu$ \\ \hline
 ABS Loop Period (Deadline) & $5000\mu$\\ \hline
\end{tabular}
\end{table}
\section{Limitations and Future Work}
\label{sec:limitation}

The main limitation of the \system is related to the low detection performances when corner cases are encountered. 
The first corner case is due to low natural frequency values: the proposed approach exploits the control theory to compute an upper and lower bound used to evaluate anomalies. These bounds are computed using the natural frequency of a second-order approximation of the control transfer function, and their steepness increases with higher natural frequency values. Consequently, lower natural frequencies may correspond to slower detection. Therefore, when designing a controller secured by \system, it is important to carefully evaluate if the detection time is compatible with the requirements of the system in terms of response to an anomaly. 
The second corner case is due to behaviors under attack that are close to the behavior of the nominal signal. In these instances, the detection time increases since the attacker remains as much as possible inside the boundaries of detection. However, as mentioned in~\ref{ssec:ov}, as long as the output value of the controller is within the bounds, the system evolves according to its expected behavior. Therefore, an attack is disruptive only once it exceeds the bounds, which triggers the detection by \system. 

Future works should explore the application of the \system approach to other \acfp{CPS}, considering the different requirements of these new domains, and potentially new, ad-hoc boundary designs.
Moreover, in the current implementation, \system detects an anomaly when the controller returns a value outside the semantic check bounds computed using the control system theory. Future works may explore the feasibility to apply machine learning for bound computation. 
To implement this in an efficient way, a solution is to use new microneural processing units, designed to be used in combination with the SOC in area-constrained embedded devices, similar to the ones considered in this work. An example of this is the new ARM Ethos-U55~\footnote{https://developer.arm.com/Processors/Ethos-U55} that, when paired with the Cortex-M55 processor, provides a 480x uplift in \ac{ML} performance over previous generation Cortex-M processors. 
\section{Conclusions}
\label{sec:tzconclusion}

In this paper, we presented \system, a new methodology that leverages \acfp{TEE} to secure real-time closed-loop controllers detecting any anomalies that would compromise the safety and performance of the \ac{CPS}. In particular, we used control theory to develop a set of bounds that model the expected system output. The proposed architecture is general, is designed to be compliant with most control algorithms, and it directly makes use  of  the \acf{TEE}  technology  of  the  micro-controller without  requiring  additional  hardware.  

We  demonstrated  the  practical  feasibility  of  our  approach  by  implementing  \system on ARM Trust-Zone and studying its performance under real-time constraints.
As a real-world scenario test case, we consider an automotive \ac{ABS}, one of the most common safety measures found in vehicles, assessing extremely promising results, proving our system's capabilities in effectively limiting the attacker. Moreover, the introduced computational and timing overheads are negligible even in relation to the heavy requirements of a real-time system such as the one considered.

\IEEEpeerreviewmaketitle

\bibliographystyle{IEEEtran}
\bibliography{bibliography}

\begin{thebibliography}{10}
\providecommand{\url}[1]{#1}
\csname url@samestyle\endcsname
\providecommand{\newblock}{\relax}
\providecommand{\bibinfo}[2]{#2}
\providecommand{\BIBentrySTDinterwordspacing}{\spaceskip=0pt\relax}
\providecommand{\BIBentryALTinterwordstretchfactor}{4}
\providecommand{\BIBentryALTinterwordspacing}{\spaceskip=\fontdimen2\font plus
\BIBentryALTinterwordstretchfactor\fontdimen3\font minus
  \fontdimen4\font\relax}
\providecommand{\BIBforeignlanguage}[2]{{%
\expandafter\ifx\csname l@#1\endcsname\relax
\typeout{** WARNING: IEEEtran.bst: No hyphenation pattern has been}%
\typeout{** loaded for the language `#1'. Using the pattern for}%
\typeout{** the default language instead.}%
\else
\language=\csname l@#1\endcsname
\fi
#2}}
\providecommand{\BIBdecl}{\relax}
\BIBdecl

\bibitem{cardenas2011attacks}
\BIBentryALTinterwordspacing
A.~A. C{\'{a}}rdenas, S.~Amin, Z.~Lin, Y.~Huang, C.~Huang, and S.~Sastry,
  ``Attacks against process control systems: risk assessment, detection, and
  response,'' in \emph{Proceedings of the 6th {ACM} Symposium on Information,
  Computer and Communications Security, {ASIACCS} 2011, Hong Kong, China, March
  22-24, 2011}, B.~S.~N. Cheung, L.~C.~K. Hui, R.~S. Sandhu, and D.~S. Wong,
  Eds.\hskip 1em plus 0.5em minus 0.4em\relax {ACM}, 2011, pp. 355--366.
  [Online]. Available: \url{https://dl.acm.org/citation.cfm?id=1966959}
\BIBentrySTDinterwordspacing

\bibitem{DBLP:conf/ccs/TronLCPZ22}
\BIBentryALTinterwordspacing
A.~de~Faveri~Tron, S.~Longari, M.~Carminati, M.~Polino, and S.~Zanero,
  ``Canflict: Exploiting peripheral conflicts for data-link layer attacks on
  automotive networks,'' in \emph{Proceedings of the 2022 {ACM} {SIGSAC}
  Conference on Computer and Communications Security, {CCS} 2022, Los Angeles,
  CA, USA, November 7-11, 2022}, H.~Yin, A.~Stavrou, C.~Cremers, and E.~Shi,
  Eds.\hskip 1em plus 0.5em minus 0.4em\relax {ACM}, 2022, pp. 711--723.
  [Online]. Available: \url{https://doi.org/10.1145/3548606.3560618}
\BIBentrySTDinterwordspacing

\bibitem{quarta2017experimental1}
\BIBentryALTinterwordspacing
D.~Quarta, M.~Pogliani, M.~Polino, F.~Maggi, A.~M. Zanchettin, and S.~Zanero,
  ``An experimental security analysis of an industrial robot controller,'' in
  \emph{2017 {IEEE} Symposium on Security and Privacy, {SP} 2017, San Jose, CA,
  USA, May 22-26, 2017}.\hskip 1em plus 0.5em minus 0.4em\relax {IEEE} Computer
  Society, 2017, pp. 268--286. [Online]. Available:
  \url{https://doi.org/10.1109/SP.2017.20}
\BIBentrySTDinterwordspacing

\bibitem{miller2015remote}
C.~Miller and C.~Valasek, ``Remote exploitation of an unaltered passenger
  vehicle,'' \emph{Black Hat USA}, vol. 2015, p.~91, 2015.

\bibitem{ding2018survey}
\BIBentryALTinterwordspacing
D.~Ding, Q.~Han, Y.~Xiang, X.~Ge, and X.~Zhang, ``A survey on security control
  and attack detection for industrial cyber-physical systems,''
  \emph{Neurocomputing}, vol. 275, pp. 1674--1683, 2018. [Online]. Available:
  \url{https://doi.org/10.1016/j.neucom.2017.10.009}
\BIBentrySTDinterwordspacing

\bibitem{hassan2019current}
\BIBentryALTinterwordspacing
M.~binti Mohamad~Noor and W.~H. Hassan, ``Current research on internet of
  things ({IoT}) security: {A} survey,'' \emph{Comput. Networks}, vol. 148, pp.
  283--294, 2019. [Online]. Available:
  \url{https://doi.org/10.1016/j.comnet.2018.11.025}
\BIBentrySTDinterwordspacing

\bibitem{longari2019secure}
\BIBentryALTinterwordspacing
S.~Longari, A.~Cannizzo, M.~Carminati, and S.~Zanero, ``A secure-by-design
  framework for automotive on-board network risk analysis,'' in \emph{2019
  {IEEE} Vehicular Networking Conference, {VNC} 2019, Los Angeles, CA, USA,
  December 4-6, 2019}.\hskip 1em plus 0.5em minus 0.4em\relax {IEEE}, 2019, pp.
  1--8. [Online]. Available:
  \url{https://doi.org/10.1109/VNC48660.2019.9062783}
\BIBentrySTDinterwordspacing

\bibitem{hasan2019protecting}
\BIBentryALTinterwordspacing
M.~Hasan and S.~Mohan, ``Protecting actuators in safety-critical {IoT} systems
  from control spoofing attacks,'' in \emph{Proceedings of the 2nd
  International {ACM} Workshop on Security and Privacy for the
  Internet-of-Things, {IoT} S{\&}P@CCS 2019, London, UK, November 15, 2019},
  P.~Liu and Y.~Zhang, Eds.\hskip 1em plus 0.5em minus 0.4em\relax {ACM}, 2019,
  pp. 8--14. [Online]. Available: \url{https://doi.org/10.1145/3338507.3358615}
\BIBentrySTDinterwordspacing

\bibitem{kim2018securing}
\BIBentryALTinterwordspacing
C.~H. Kim, T.~Kim, H.~Choi, Z.~Gu, B.~Lee, X.~Zhang, and D.~Xu, ``Securing
  real-time microcontroller systems through customized memory view switching,''
  in \emph{25th Annual Network and Distributed System Security Symposium,
  {NDSS} 2018, San Diego, California, USA, February 18-21, 2018}.\hskip 1em
  plus 0.5em minus 0.4em\relax The Internet Society, 2018. [Online]. Available:
  \url{http://wp.internetsociety.org/ndss/wp-content/uploads/sites/25/2018/02/ndss2018\_04B-2\_Kim\_paper.pdf}
\BIBentrySTDinterwordspacing

\bibitem{visioli2006practical}
\BIBentryALTinterwordspacing
M.~Beschi, A.~Piazzi, and A.~Visioli, ``On the practical implementation of a
  noncausal feedforward technique for {PID} control,'' in \emph{10th European
  Control Conference, {ECC} 2009, Budapest, Hungary, 23-26 August 2009}.\hskip
  1em plus 0.5em minus 0.4em\relax {IEEE}, 2009, pp. 1806--1811. [Online].
  Available: \url{https://doi.org/10.23919/ECC.2009.7074665}
\BIBentrySTDinterwordspacing

\bibitem{aastrom2013computer}
\BIBentryALTinterwordspacing
C.~R.~J. Jr., ``Computer-controlled systems: Theory and design : Karl j.
  {\aa}str{\"{o}}m and bj{\"{o}}rn wittenmark,'' \emph{Autom.}, vol.~21, no.~6,
  pp. 746--748, 1985. [Online]. Available:
  \url{https://doi.org/10.1016/0005-1098(85)90050-0}
\BIBentrySTDinterwordspacing

\bibitem{gul2020review}
F.~Gul, S.~S.~N. Alhady, and W.~Rahiman, ``A review of controller approach for
  autonomous guided vehicle system,'' \emph{Indonesian Journal of Electrical
  Engineering and Computer Science}, vol.~20, no.~1, pp. 552--562, 2020.

\bibitem{abc}
S.~M. Savaresi and M.~Tanelli, \emph{Active braking control systems design for
  vehicles}.\hskip 1em plus 0.5em minus 0.4em\relax Springer Science \&
  Business Media, 2010.

\bibitem{holding2009arm}
A.~Holding, ``Arm security technology: Building a secure system using trustzone
  technology,'' 2009.

\bibitem{finnicum2011building}
\BIBentryALTinterwordspacing
M.~Finnicum and S.~T. King, ``Building secure robot applications,'' in
  \emph{6th {USENIX} Workshop on Hot Topics in Security, HotSec'11, San
  Francisco, CA, USA, August 9, 2011}, P.~D. McDaniel, Ed.\hskip 1em plus 0.5em
  minus 0.4em\relax {USENIX} Association, 2011. [Online]. Available:
  \url{https://www.usenix.org/conference/hotsec11/building-secure-robot-applications}
\BIBentrySTDinterwordspacing

\bibitem{amin2013cyber}
\BIBentryALTinterwordspacing
S.~Amin, X.~Litrico, S.~Sastry, and A.~M. Bayen, ``Cyber security of water
  {SCADA} systems - part {I:} analysis and experimentation of stealthy
  deception attacks,'' \emph{{IEEE} Trans. Control. Syst. Technol.}, vol.~21,
  no.~5, pp. 1963--1970, 2013. [Online]. Available:
  \url{https://doi.org/10.1109/TCST.2012.2211873}
\BIBentrySTDinterwordspacing

\bibitem{liu2009false}
\BIBentryALTinterwordspacing
Y.~Liu, M.~K. Reiter, and P.~Ning, ``False data injection attacks against state
  estimation in electric power grids,'' in \emph{Proceedings of the 2009 {ACM}
  Conference on Computer and Communications Security, {CCS} 2009, Chicago,
  Illinois, USA, November 9-13, 2009}, E.~Al{-}Shaer, S.~Jha, and A.~D.
  Keromytis, Eds.\hskip 1em plus 0.5em minus 0.4em\relax {ACM}, 2009, pp.
  21--32. [Online]. Available: \url{https://doi.org/10.1145/1653662.1653666}
\BIBentrySTDinterwordspacing

\bibitem{bhardwaj2019cyber}
\BIBentryALTinterwordspacing
A.~Bhardwaj, V.~Avasthi, and S.~Goundar, ``Cyber security attacks on robotic
  platforms,'' \emph{Netw. Secur.}, vol. 2019, no.~10, pp. 13--19, 2019.
  [Online]. Available: \url{https://doi.org/10.1016/S1353-4858(19)30122-9}
\BIBentrySTDinterwordspacing

\bibitem{goh2017anomaly}
\BIBentryALTinterwordspacing
J.~Goh, S.~Adepu, M.~Tan, and Z.~S. Lee, ``Anomaly detection in cyber physical
  systems using recurrent neural networks,'' in \emph{18th {IEEE} International
  Symposium on High Assurance Systems Engineering, {HASE} 2017, Singapore,
  January 12-14, 2017}.\hskip 1em plus 0.5em minus 0.4em\relax {IEEE} Computer
  Society, 2017, pp. 140--145. [Online]. Available:
  \url{https://doi.org/10.1109/HASE.2017.36}
\BIBentrySTDinterwordspacing

\bibitem{giraldo2018survey}
\BIBentryALTinterwordspacing
J.~Giraldo, D.~I. Urbina, A.~A. C{\'{a}}rdenas, J.~Valente, M.~A. Faisal,
  J.~Ruths, N.~O. Tippenhauer, H.~Sandberg, and R.~Candell, ``A survey of
  physics-based attack detection in cyber-physical systems,'' \emph{{ACM}
  Comput. Surv.}, vol.~51, no.~4, pp. 76:1--76:36, 2018. [Online]. Available:
  \url{https://doi.org/10.1145/3203245}
\BIBentrySTDinterwordspacing

\bibitem{hadziosmanovic2014through}
\BIBentryALTinterwordspacing
D.~Hadziosmanovic, R.~Sommer, E.~Zambon, and P.~H. Hartel, ``Through the eye of
  the {PLC:} semantic security monitoring for industrial processes,'' in
  \emph{Proceedings of the 30th Annual Computer Security Applications
  Conference, {ACSAC} 2014, New Orleans, LA, USA, December 8-12, 2014},
  C.~N.~P. Jr., A.~Hahn, K.~R.~B. Butler, and M.~Sherr, Eds.\hskip 1em plus
  0.5em minus 0.4em\relax {ACM}, 2014, pp. 126--135. [Online]. Available:
  \url{https://doi.org/10.1145/2664243.2664277}
\BIBentrySTDinterwordspacing

\bibitem{mashima2012research}
\BIBentryALTinterwordspacing
D.~Mashima and A.~A. C{\'{a}}rdenas, ``Evaluating electricity theft detectors
  in smart grid networks,'' in \emph{Research in Attacks, Intrusions, and
  Defenses - 15th International Symposium, {RAID} 2012, Amsterdam, The
  Netherlands, September 12-14, 2012. Proceedings}, ser. Lecture Notes in
  Computer Science, D.~Balzarotti, S.~J. Stolfo, and M.~Cova, Eds., vol.
  7462.\hskip 1em plus 0.5em minus 0.4em\relax Springer, 2012, pp. 210--229.
  [Online]. Available: \url{https://doi.org/10.1007/978-3-642-33338-5\_11}
\BIBentrySTDinterwordspacing

\bibitem{wang2014srid}
\BIBentryALTinterwordspacing
Y.~Wang, Z.~Xu, J.~Zhang, L.~Xu, H.~Wang, and G.~Gu, ``{SRID:} state relation
  based intrusion detection for false data injection attacks in {SCADA},'' in
  \emph{Computer Security - {ESORICS} 2014 - 19th European Symposium on
  Research in Computer Security, Wroclaw, Poland, September 7-11, 2014.
  Proceedings, Part {II}}, ser. Lecture Notes in Computer Science,
  M.~Kutylowski and J.~Vaidya, Eds., vol. 8713.\hskip 1em plus 0.5em minus
  0.4em\relax Springer, 2014, pp. 401--418. [Online]. Available:
  \url{https://doi.org/10.1007/978-3-319-11212-1\_23}
\BIBentrySTDinterwordspacing

\bibitem{chong2015observability}
\BIBentryALTinterwordspacing
M.~S. Chong, M.~Wakaiki, and J.~P. Hespanha, ``Observability of linear systems
  under adversarial attacks,'' in \emph{American Control Conference, {ACC}
  2015, Chicago, IL, USA, July 1-3, 2015}.\hskip 1em plus 0.5em minus
  0.4em\relax {IEEE}, 2015, pp. 2439--2444. [Online]. Available:
  \url{https://doi.org/10.1109/ACC.2015.7171098}
\BIBentrySTDinterwordspacing

\bibitem{shoukry2020smt}
\BIBentryALTinterwordspacing
Y.~Shoukry, M.~Chong, M.~Wakaiki, P.~Nuzzo, A.~L. Sangiovanni{-}Vincentelli,
  S.~A. Seshia, J.~P. Hespanha, and P.~Tabuada, ``Smt-based observer design for
  cyber-physical systems under sensor attacks,'' \emph{{ACM} Trans. Cyber Phys.
  Syst.}, vol.~2, no.~1, pp. 5:1--5:27, 2018. [Online]. Available:
  \url{https://doi.org/10.1145/3078621}
\BIBentrySTDinterwordspacing

\bibitem{mo2009secure}
Y.~Mo and B.~Sinopoli, ``Secure control against replay attacks,'' in \emph{2009
  47th annual Allerton conference on communication, control, and computing
  (Allerton)}.\hskip 1em plus 0.5em minus 0.4em\relax IEEE, 2009, pp. 911--918.

\bibitem{mo2015physical}
Y.~Mo, S.~Weerakkody, and B.~Sinopoli, ``Physical authentication of control
  systems: Designing watermarked control inputs to detect counterfeit sensor
  outputs,'' \emph{IEEE Control Systems Magazine}, vol.~35, no.~1, pp. 93--109,
  2015.

\bibitem{valente2015using}
\BIBentryALTinterwordspacing
J.~Valente and A.~A. C{\'{a}}rdenas, ``Using visual challenges to verify the
  integrity of security cameras,'' in \emph{Proceedings of the 31st Annual
  Computer Security Applications Conference, Los Angeles, CA, USA, December
  7-11, 2015}.\hskip 1em plus 0.5em minus 0.4em\relax {ACM}, 2015, pp.
  141--150. [Online]. Available: \url{https://doi.org/10.1145/2818000.2818045}
\BIBentrySTDinterwordspacing

\bibitem{pu2020detecting}
\BIBentryALTinterwordspacing
H.~Pu, L.~He, C.~Zhao, D.~K.~Y. Yau, P.~Cheng, and J.~Chen, ``Detecting replay
  attacks against industrial robots via power fingerprinting,'' in \emph{SenSys
  '20: The 18th {ACM} Conference on Embedded Networked Sensor Systems, Virtual
  Event, Japan, November 16-19, 2020}, J.~Nakazawa and P.~Huang, Eds.\hskip 1em
  plus 0.5em minus 0.4em\relax {ACM}, 2020, pp. 285--297. [Online]. Available:
  \url{https://doi.org/10.1145/3384419.3430775}
\BIBentrySTDinterwordspacing

\bibitem{kiss2015clustering}
\BIBentryALTinterwordspacing
I.~Kiss, B.~Genge, and P.~Haller, ``A clustering-based approach to detect cyber
  attacks in process control systems,'' in \emph{13th {IEEE} International
  Conference on Industrial Informatics, {INDIN} 2015, Cambridge, United
  Kingdom, July 22-24, 2015}.\hskip 1em plus 0.5em minus 0.4em\relax {IEEE},
  2015, pp. 142--148. [Online]. Available:
  \url{https://doi.org/10.1109/INDIN.2015.7281725}
\BIBentrySTDinterwordspacing

\bibitem{hei2013pipac}
\BIBentryALTinterwordspacing
X.~Hei, X.~Du, S.~Lin, and I.~Lee, ``{PIPAC:} patient infusion pattern based
  access control scheme for wireless insulin pump system,'' in
  \emph{Proceedings of the {IEEE} {INFOCOM} 2013, Turin, Italy, April 14-19,
  2013}.\hskip 1em plus 0.5em minus 0.4em\relax {IEEE}, 2013, pp. 3030--3038.
  [Online]. Available: \url{https://doi.org/10.1109/INFCOM.2013.6567115}
\BIBentrySTDinterwordspacing

\bibitem{longari2020cannolo}
\BIBentryALTinterwordspacing
S.~Longari, D.~H.~N. Valcarcel, M.~Zago, M.~Carminati, and S.~Zanero,
  ``Cannolo: An anomaly detection system based on {LSTM} autoencoders for
  controller area network,'' \emph{{IEEE} Trans. Netw. Serv. Manag.}, vol.~18,
  no.~2, pp. 1913--1924, 2021. [Online]. Available:
  \url{https://doi.org/10.1109/TNSM.2020.3038991}
\BIBentrySTDinterwordspacing

\bibitem{altawy2016security}
\BIBentryALTinterwordspacing
R.~AlTawy and A.~M. Youssef, ``Security tradeoffs in cyber physical systems:
  {A} case study survey on implantable medical devices,'' \emph{{IEEE} Access},
  vol.~4, pp. 959--979, 2016. [Online]. Available:
  \url{https://doi.org/10.1109/ACCESS.2016.2521727}
\BIBentrySTDinterwordspacing

\bibitem{komninos2014survey}
\BIBentryALTinterwordspacing
N.~Komninos, E.~Philippou, and A.~Pitsillides, ``Survey in smart grid and smart
  home security: Issues, challenges and countermeasures,'' \emph{{IEEE} Commun.
  Surv. Tutorials}, vol.~16, no.~4, pp. 1933--1954, 2014. [Online]. Available:
  \url{https://doi.org/10.1109/COMST.2014.2320093}
\BIBentrySTDinterwordspacing

\bibitem{DBLP:journals/mam/YaacoubSNKCM20}
\BIBentryALTinterwordspacing
J.~A. Yaacoub, O.~Salman, H.~N. Noura, N.~Kaaniche, A.~Chehab, and M.~Malli,
  ``Cyber-physical systems security: Limitations, issues and future trends,''
  \emph{Microprocess. Microsystems}, vol.~77, p. 103201, 2020. [Online].
  Available: \url{https://doi.org/10.1016/j.micpro.2020.103201}
\BIBentrySTDinterwordspacing

\bibitem{liu2017protc}
\BIBentryALTinterwordspacing
R.~Liu and M.~B. Srivastava, ``{PROTC:} protecting drone's peripherals through
  {ARM} trustzone,'' in \emph{Proceedings of the 3rd Workshop on Micro Aerial
  Vehicle Networks, Systems, and Applications, DroNet@MobiSys 2017, Niagara
  Falls, NY, USA, June 23, 2017}.\hskip 1em plus 0.5em minus 0.4em\relax {ACM},
  2017, pp. 1--6. [Online]. Available:
  \url{https://doi.org/10.1145/3086439.3086443}
\BIBentrySTDinterwordspacing

\bibitem{yoon2017virtualdrone}
\BIBentryALTinterwordspacing
M.~Yoon, B.~Liu, N.~Hovakimyan, and L.~Sha, ``Virtualdrone: virtual sensing,
  actuation, and communication for attack-resilient unmanned aerial systems,''
  in \emph{Proceedings of the 8th International Conference on Cyber-Physical
  Systems, {ICCPS} 2017, Pittsburgh, Pennsylvania, USA, April 18-20, 2017},
  S.~Mart{\'{\i}}nez, E.~Tovar, C.~Gill, and B.~Sinopoli, Eds.\hskip 1em plus
  0.5em minus 0.4em\relax {ACM}, 2017, pp. 143--154. [Online]. Available:
  \url{https://doi.org/10.1145/3055004.3055010}
\BIBentrySTDinterwordspacing

\bibitem{zhang2000adaptive}
\BIBentryALTinterwordspacing
T.~Zhang, S.~S. Ge, and C.~C. Hang, ``Adaptive neural network control for
  strict-feedback nonlinear systems using backstepping design,'' \emph{Autom.},
  vol.~36, no.~12, pp. 1835--1846, 2000. [Online]. Available:
  \url{https://doi.org/10.1016/S0005-1098(00)00116-3}
\BIBentrySTDinterwordspacing

\bibitem{aastrom2010feedback}
K.~{\AA}str{\"o}m and R.~M. Murray, \emph{Feedback systems}.\hskip 1em plus
  0.5em minus 0.4em\relax Princeton university press, 2010.

\bibitem{tanelli2007combining}
\BIBentryALTinterwordspacing
M.~Tanelli, R.~Sartori, and S.~M. Savaresi, ``Combining slip and deceleration
  control for brake-by-wire control systems: {A} sliding-mode approach,''
  \emph{Eur. J. Control}, vol.~13, no.~6, pp. 593--611, 2007. [Online].
  Available: \url{https://doi.org/10.3166/ejc.13.593-611}
\BIBentrySTDinterwordspacing

\end{thebibliography}

\begin{IEEEbiography}[{\includegraphics[width=1in,height=1.25in,clip,keepaspectratio]{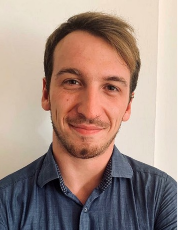}}]{Stefano Longari}
 received a Ph.D. in Information Technology from Politecnico di Milano, where he currently works as a Postdoctoral Researcher in NECST laboratory as part of the System Security group inside the Dipartimento di Elettronica, Informazione e Bioingegneria. His main research focus is automotive security and the use of novel technologies such as machine learning intrusion detection to secure the development of smart mobility and smart city ecosystems. Since 2020 he is a member of the Automotive Security Research Group.
\end{IEEEbiography}
\begin{IEEEbiography}[{\includegraphics[width=1in,height=1.25in,clip,keepaspectratio]{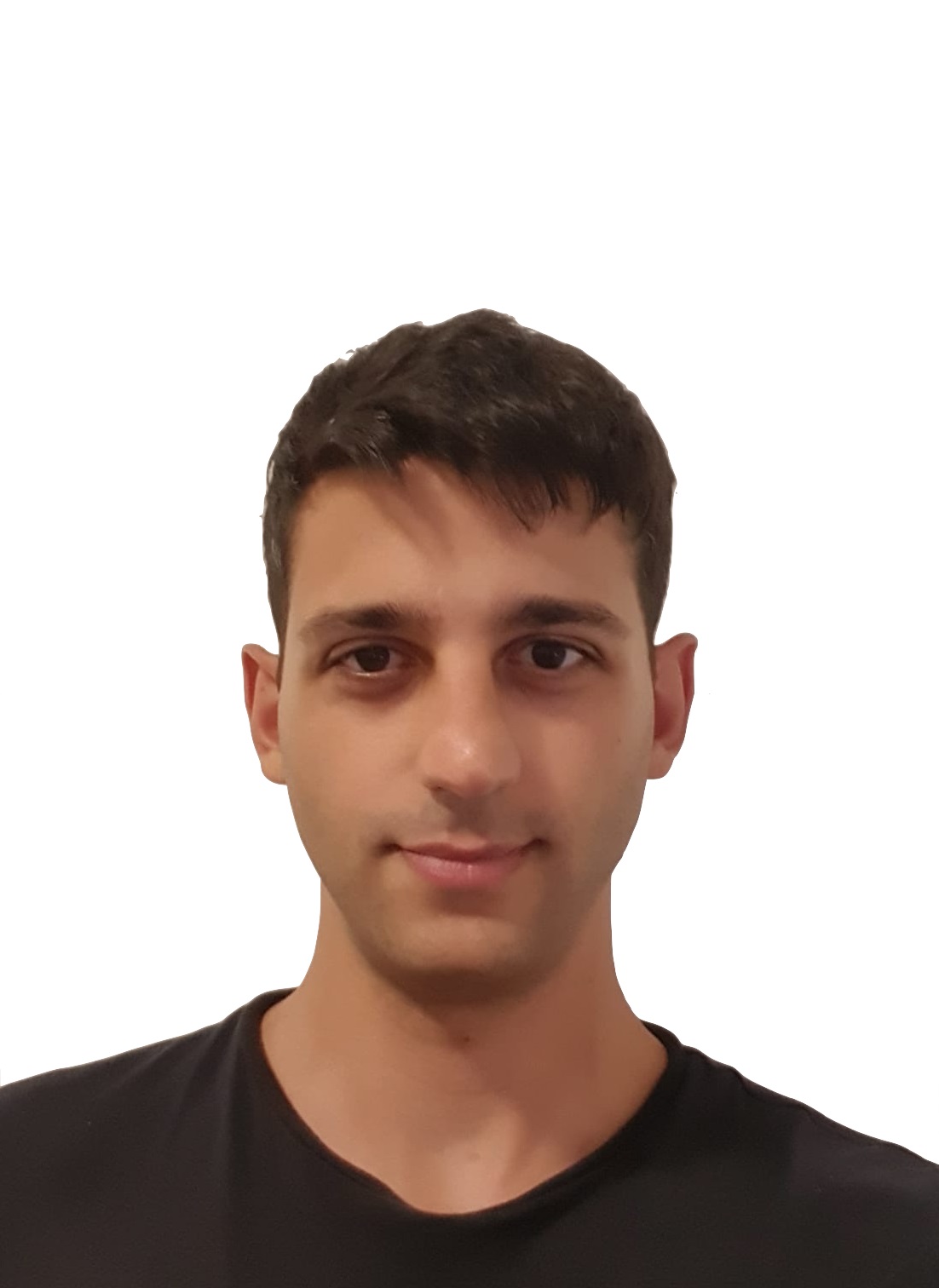}}]{Alessandro Pozone}
was born in L'Aquila, Italy. During his B.Sc. in Computer Engineering from Politecnico di Milano he worked on Reverse, a project focused on using augmented reality to show how cities developed over time with the focus on L’Aquila, a place radically changed by the 2009 earthquake. He got his M.Sc. after defending a thesis on exploiting the available security measures in ARM CPUs to secure the control systems world, mainly in the case of real time systems. He is now a Software Engineer at Microsoft.
\end{IEEEbiography}
\begin{IEEEbiography}
[{\includegraphics[width=1in,height=1.25in,clip,keepaspectratio]{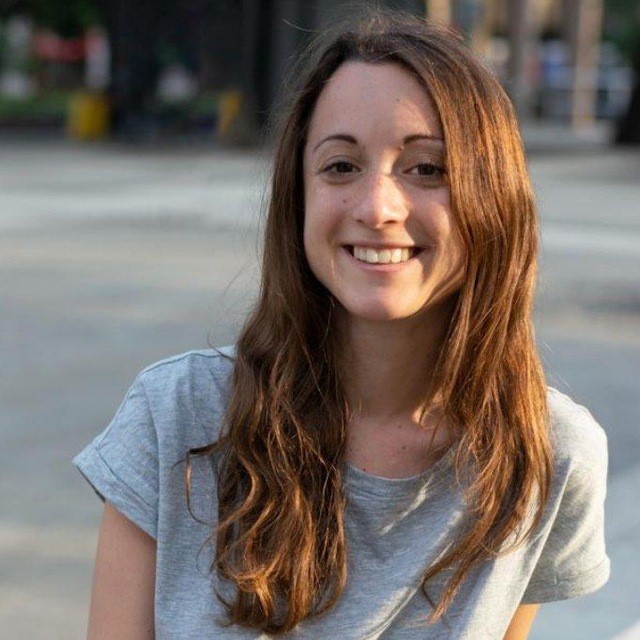}}]{Jessica Leoni} was born in Treviglio, Italy in 1995. Jessica is a researcher assistant at Politecnico di Milano, and received a Master Degree in Biomedical Engineering (cum laude) in 2019 and a Ph.D. in Data Analytics and Decision Sciences (cum laude) in 2022. 
Her research focuses on the design of safety-oriented health and usage monitoring systems for critical vehicles. Her expertise lies in advanced machine- and deep-learning techniques, which she utilizes to design innovative solutions paving the way for safer and more efficient transportation systems. In addition, she leverages her biomedical background to investigate how drivers' psychophysiological status affects their usage and the vehicle's performance.
\end{IEEEbiography}
\begin{IEEEbiography}
[{\includegraphics[width=1in,height=1.25in,clip,keepaspectratio]{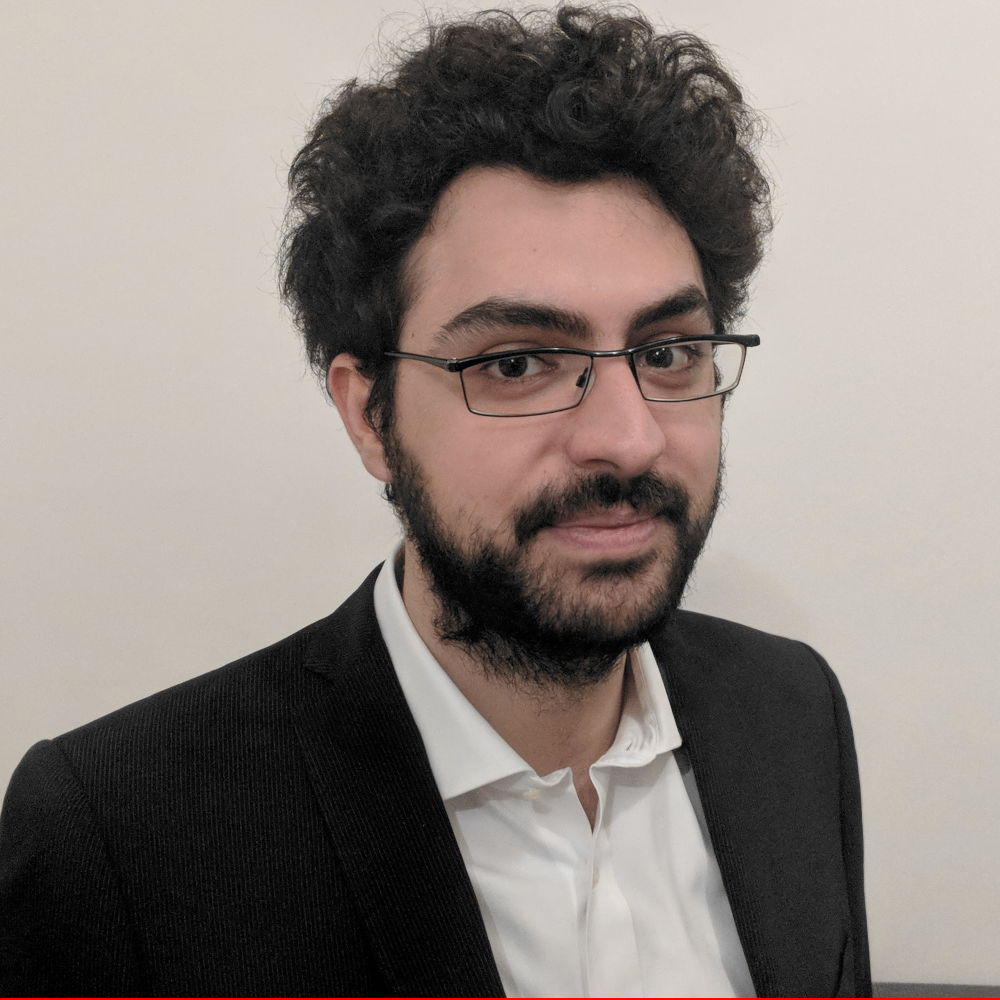}}]{Mario Polino} has got his PhD (2017) in
Computer Security from Politecnico di Milano.
Currently, Mario is an Assistant Professor at Politecnico
di Milano. His Ph.D. was mainly based on
Malware Analysis and automation of behavior extraction.
Besides malware analysis, Mario is interested
in several aspects of Computer Security; his
works range from the study of Cyber-Physical systems
to Banking Fraud Detection going through
Android Security. Even more, he is interested in
Binary analysis and Exploitation techniques. Mario spends most of his free
time playing Capture the Flag Competition with "Tower of Hanoi" and
"mhackeroni".
\end{IEEEbiography}
\begin{IEEEbiography}
[{\includegraphics[width=1in,height=1.25in,clip,keepaspectratio]{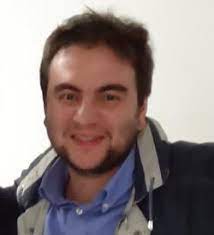}}]{Michele Carminati} received his Ph.D. degree cum laude in Information Technology from Politecnico di Milano in Italy, where he is currently an Assistant Professor working at NECST laboratory as part of the System Security group inside the Dipartimento di Elettronica, Informazione e Bioingegneria. His research revolves around the application of machine learning methods in various cybersecurity-related fields, ranging from cyber-physical and automotive systems to binary and malware analysis, going through anomaly and intrusion detection. He is actively involved in research projects funded by the European Union, and he is also co-founder of Banksealer, a Fintech spinoff of Politecnico di Milano.
\end{IEEEbiography}
\begin{IEEEbiography}
[{\includegraphics[width=1in,height=1.25in,clip,keepaspectratio]{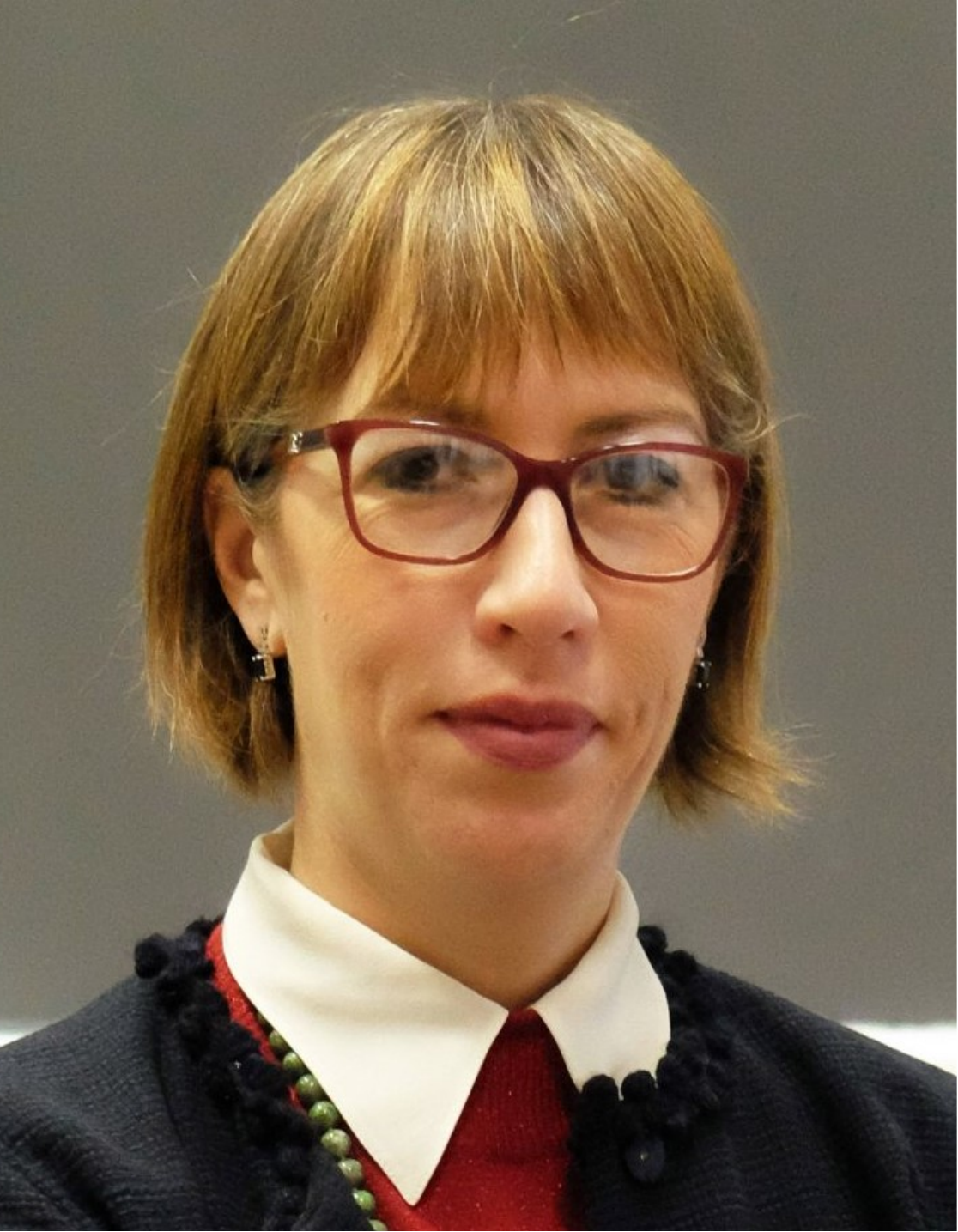}}]{Mara Tanelli}
(M'05, SM'12) is currently a Full Professor of Automatic Control at the Politecnico di Milano, where she earned her Ph.D. in Information Engineering with honors in 2007. Her main research interests are in Automotive Control, Smart Mobility and Industrial Analytics. She is co-author of more than 170 peer-reviewed publications in these research areas. Prof. Tanelli is a member of the Conference Editorial Board of the IEEE Control Systems Society (CSS). She is AE for the IEEE Transactions on Human-Machine Systems. Since 2021 she is Chair of the Technical Committee (TC) on Automotive Controls of the IEEE CSS and vice-chair for publications of the corresponding IFAC TC.
\end{IEEEbiography}
\begin{IEEEbiography}
[{\includegraphics[width=1in,height=1.25in,clip,keepaspectratio]{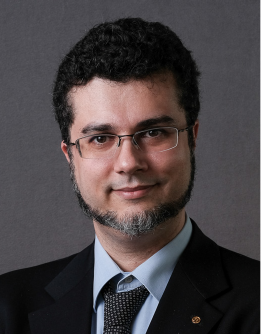}}]{Stefano Zanero}
(M’03,  SM’10) received a Ph.D. degree in Computer Engineering from
Politecnico di Milano, where he is currently a Full Professor with
the Dipartimento di Elettronica, Informazione e Bioingegneria. His
research focuses on cybersecurity, in particular, cyber-physical systems,
malware analysis, and fraud and anomaly detection through the use of
machine learning. Prof. Zanero is a Distinguished Contributor of the IEEE Computer Society, where he currently serves on the Board of Governors.
\end{IEEEbiography}
\end{document}